\documentstyle[11pt,newpasp,twoside,epsf]{article}
\begin{document}
\title{H1504+65 -- The Naked Stellar C/O Core of a Former Red Giant Observed
with FUSE and Chandra}
 \author{Klaus Werner, Thomas Rauch}
\affil{Institut f\"ur Astronomie und Astrophysik, Univ.\ T\"ubingen, Germany}
\author{Martin A. Barstow}
\affil{Department of Physics and Astronomy, University of Leicester, UK}
\author{Jeff W. Kruk}
\affil{Department of Physics and Astronomy, JHU, Baltimore MD, U.S.A.}

\begin{abstract}
\ H1504+65 is an extremely hot hydrogen-deficient white dwarf with an effective
temperature close to 200\,000\,K. We present new FUV and soft X-ray spectra
obtained with FUSE and Chandra, which confirm that H1504+65 has an atmosphere
primarily composed of carbon and oxygen. The Chandra spectra show a wealth of
absorption lines from highly ionized oxygen, neon and magnesium and suggest
relatively high Ne and Mg abundances. This corroborates an earlier suggestion
that H1504+65 represents a naked C/O stellar core or even the C/O envelope of a
O-Ne-Mg white dwarf.
\end{abstract}

\section{Introduction}

H1504+65 is an optically faint blue star. It has been identified as the
counterpart of a bright soft X-ray source (Nousek et al.\ 1986) that was
discovered by an early X-ray survey (Nugent et al.\ 1983). Spectroscopically,
the star is a member of the PG1159 class, which comprises hot hydrogen-deficient
post-AGB stars ($T_{\rm eff}$=75\,000\,K--180\,000\,K, $\log g$=5.5--8; Werner
2001). The PG1159 stars are probably the outcome of a late helium-shell flash,
a phenomenon that drives the currently observed fast evolutionary rates of
three well-know objects (FG~Sge, Sakurai's object, V605 Aql). A late
helium-shell flash may occur in a post-AGB star or a white dwarf, and brings
the star back onto the AGB. Flash induced mixing generates a H-deficient surface
layer and the subsequent second post-AGB evolution explains the existence of
Wolf-Rayet central stars of planetary nebulae and their successors, the PG1159
stars (see also the review of Sch\"onberner \& Jeffery in these proceedings).
Within the PG1159 group H1504+65 is an extraordinary object, as it has been
shown that H1504+65 is not only hydrogen-deficient but also helium-deficient.
From optical spectra it was concluded that the atmosphere is primarily composed
of carbon and oxygen, by equal amounts (Werner 1991). Neon lines were detected
in soft X-ray spectra taken with the EUVE satellite and in an optical-UV Keck
spectrum and an abundance of Ne=2--5\% (all abundances in this paper are given
as mass fractions) was derived (Werner \& Wolff 1999).

We present here first results of new observations performed with FUSE and
Chandra, whose spectroscopic resolution is more than an order of magnitude
better than that of previous FUV and EUV missions (HUT and EUVE). Synthetic
spectra were computed with a NLTE code (Werner \& Dreizler 1999) which
constructs line blanketed, static, plane parallel model atmospheres.

\begin{figure}
\epsfxsize=\textwidth
\epsffile{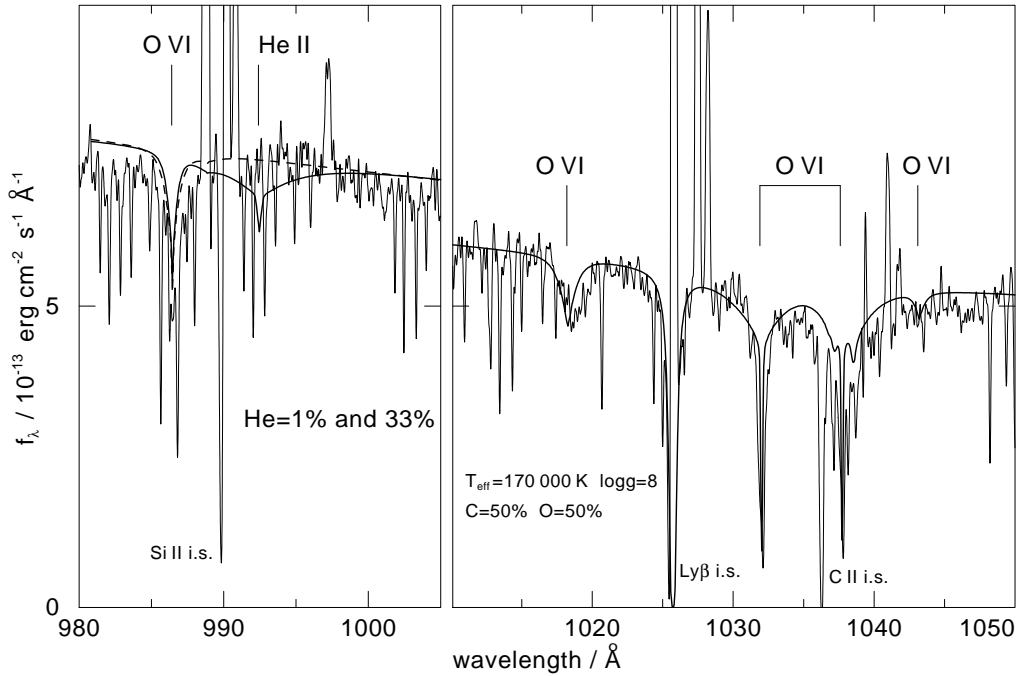}
\caption{
Details from the FUSE spectrum of H1504+65. Left panel: The lack of the He\,II
(n=2$\rightarrow$7) absorption line confirms the He-deficiency. Overplotted are
two models with He=33\% and He=1\%. Right panel: The O\,VI resonance doublet
shows no signature of a stellar wind. Narrow absorption lines are of
interstellar origin (mostly H$_2$) and emission lines are from the geocorona.
All spectra are smoothed with Gaussians (0.1\AA\ FWHM).
}\label{fuse}
\end{figure}

\section{FUSE and Chandra spectroscopy}

H1504+65 was observed with FUSE on Jan.\ 28, 2002, with an integration time of
10.7 hours. The spectral range covered was 912\AA--1188\AA\ with a resolution of
about 0.1\AA. The spectrum is characterized by a few broad photospheric lines
of C\,IV and O\,VI and many narrow interstellar lines, predominantly from
H$_2$. Two interesting details are shown in Fig.\,\ref{fuse}. The left panel is
centered around the location of a He\,II line. It is clearly seen that no
He\,II absorption is detectable and a comparison with models shows that the
helium abundance is at most of the order of 1\%. This confirms the results from
a less reliable optical analysis, which was entirely based on the lack of a
He\,II~4686\AA\ emission line (Werner 1991), and from HUT spectroscopy (Kruk \&
Werner 1998). The right panel is centered around the O\,VI resonance doublet.
The observed profile shows no evidence for on-going mass-loss and hence can be
fitted with a synthetic profile of a static model. This will allow the
derivation of an upper limit for the mass-loss rate from which one can
determine the significance of mass-loss for the cooling rate and the surface
chemical composition.

\begin{figure}
\epsfxsize=\textwidth
\epsffile{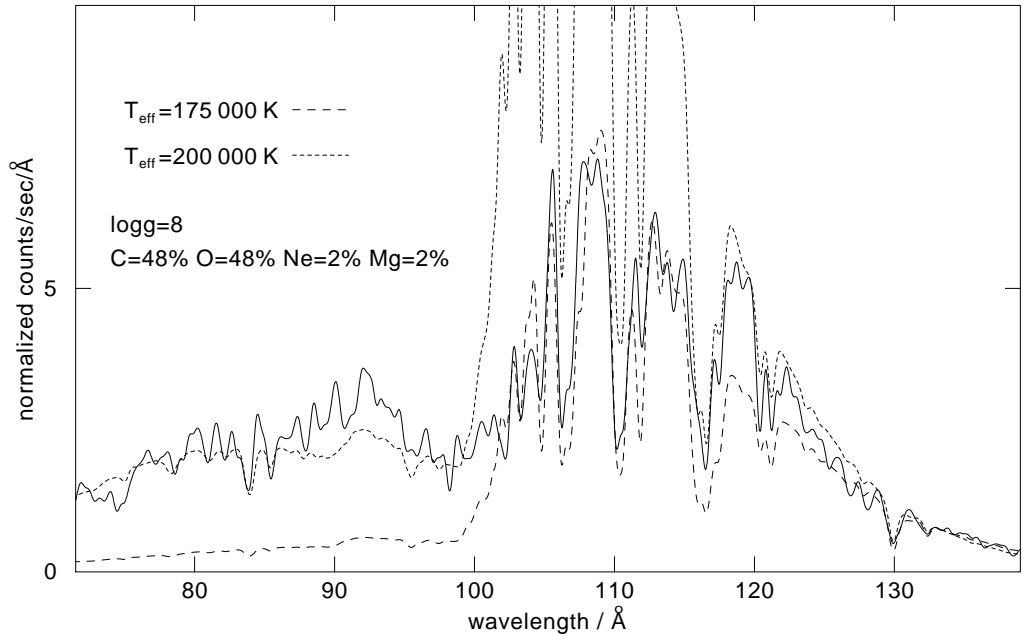}
\caption{
Overview of the Chandra spectrum of H1504+65 (thick line). Two models with
different $T_{\rm eff}$ are shown. The 175\,000\,K model fits the overall flux
at long wavelengths, while it underestimates the short wavelength flux (dashed
line). The hotter model (200\,000\,K, solid line) fits at short wavelengths but
overestimates the long wavelength flux. The model spectra were attenuated by an
ISM model with $n_{\rm H}=8.2 \cdot 10^{19}$\,cm$^{-2}$ and $9.4 \cdot
10^{19}$\,cm$^{-2}$, respectively, then folded through the instrument response
and normalized to the observation to fit near 130\AA. For clarity, all spectra
are smoothed with Gaussians (0.5\AA\ FWHM).
}\label{Chandra}
\end{figure}

\begin{figure}
[ht]
\epsfxsize=\textwidth
\epsffile{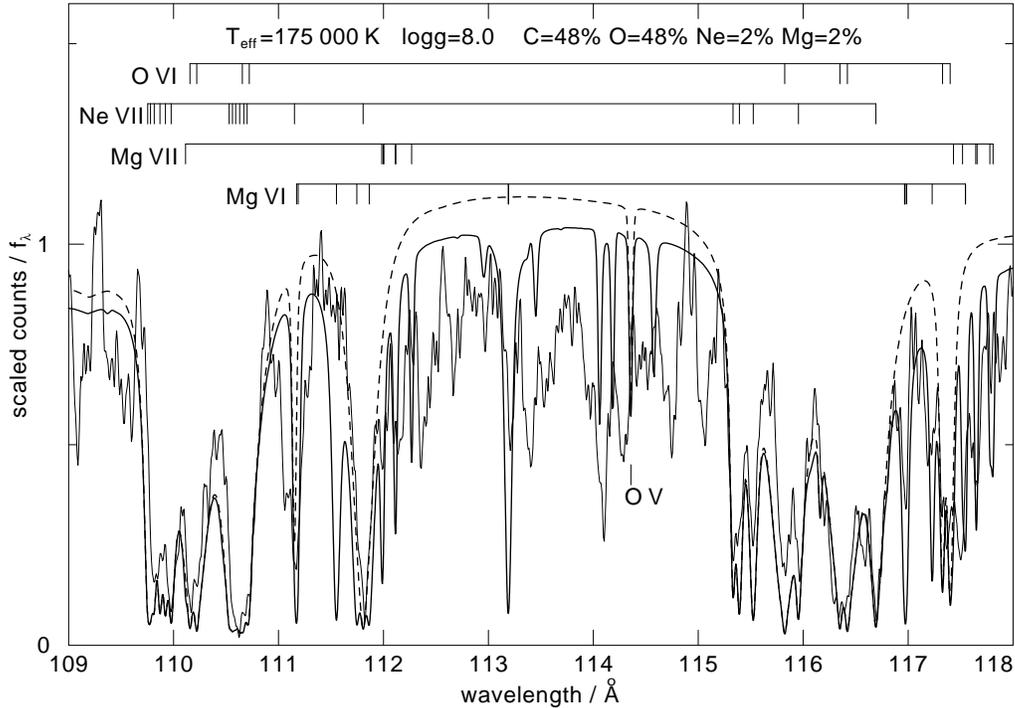}
\caption{
Detail from the Chandra count spectrum of H1504+65 (thin line), shifted by
$-0.06$\AA\ to zero radial velocity.  Relative fluxes of two models are
overplotted, one including Mg (solid line) and one without Mg (dashed line), in
order to facilitate the identification of Mg lines. Observation and model
spectra are smoothed with Gaussians (FWHM 0.02\AA\ and 0.03\AA, respectively).
}\label{Chandra_109}
\end{figure}

\begin{figure}
[!ht]
\epsfxsize=\textwidth
\epsffile{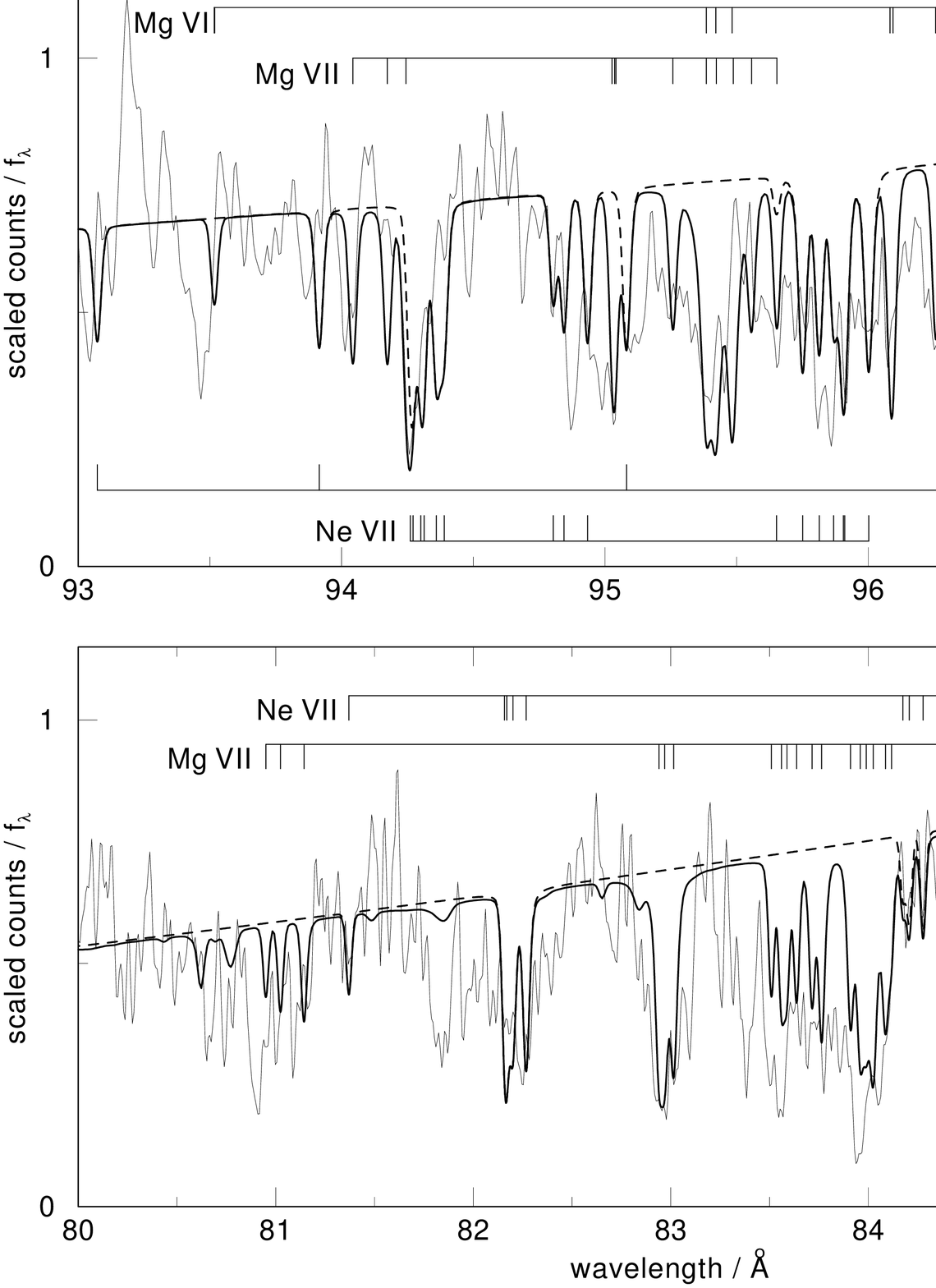}
\caption{
Like Fig.\,\ref{Chandra_109}, but other spectral regions. All lines included in
the models are identified. Additional weak absorption features in the models
are due to resonances in the bound-free cross-sections.
}\label{Chandra_80}
\end{figure}

H1504+65 was observed with Chandra on Sep.\ 27, 2000, with an integration time
of 7 hours. Flux was detected in the range 60\AA--160\AA\ and the spectral
resolution is about 0.1\AA. Fig.\,\ref{Chandra} shows the overall spectrum. It
is characterized by a roll-off at long wavelengths due to ISM absorption. The
maximum flux is detected near 110\AA. Between 105\AA\ and 100\AA\ the flux
drops because of absorption from the O\,VI edge of the first excited atomic
level. The edge is not sharp because of a converging line series and pressure
ionization (see Werner \& Wolff, 1999, for detailed model spectra). Below
100\AA\ the flux decreases, representing the Wien tail of the photospheric flux
distribution. Fig.\,\ref{Chandra} demonstrates that current models cannot fit
the entire wavelength range at a unique temperature, so that $T_{\rm eff}$ can
only be constrained between 175\,000\,K and 200\,000\,K. At present we favor
the higher value, because the excess model flux at 100\AA--120\AA\ can be
explained by missing opacities from iron group elements, as exploratory
calculations have shown. Also, a detailed inspection of the relative strengths
of Mg\,VI and Mg\,VII lines favors the higher temperature.

Figs.\,\ref{Chandra_109} and \ref{Chandra_80} display astonishing details in
the Chandra spectrum. At first sight the spectrum appears to be rather noisy,
but comparison with model spectra suggests that most absorption features in the
observation, which are not yet fitted with current models, are probably real.
Magnesium lines are discovered for the first time in H1504+65, which however
becomes only clear by comparison with modeling with a very detailed Mg model
atom (see in particular Fig.\,\ref{Chandra_80}, top panel). It comprises more
than 120 NLTE levels and 130 individual lines in the Chandra spectral range. It
is worth mentioning that fine structure splitting of the metal lines can be
detected in the observed spectra down to a separation of only 0.1\AA\ and
requires such detailed modeling. Designing the respective model atoms is
cumbersome because atomic data from different sources needs to be collected.
Energy levels are obtained from either Bashkin \& Stoner (1975) or the NIST
database. Oscillator strengths are taken from TOPBASE, the Opacity Project (OP)
database (Seaton et al.\ 1994). Whenever possible, bound-free cross-sections
for atomic levels are also taken from TOPBASE, but in some cases hydrogen-like
approximations are necessary. A problem is posed by strong resonances in the OP
photoionization data. They are responsible for some strong, absorption
line-like features in the computed spectra. Their exact wavelength location is,
however, uncertain because the data are results from ab-initio calculations
yielding inaccurate energy levels. Collisional rates are computed by usual
formulae which employ radiative cross-sections as far as possible.

As already mentioned, we have computed exploratory models including opacities
of heavy metals (Ca--Ni). This introduces a very large number of additional
lines which effectively block flux in the 100\AA--120\AA\ region and thus may
eventually lead to a unique temperature determination. However, identifying
individual heavy metal lines will be problematic, because the vast majority of
lines in the available lists (Kurucz 1994) have uncertain wavelength positions.

\section{Discussion}

From our first models presented here, the magnesium abundance is of the order
2\%, similar to the Ne abundance determined earlier. This allows two
alternative interpretations. Either we see a naked C/O white dwarf with
3$\alpha$ processed matter, which is, according to evolutionary models,
dominated by C and O with an admixture of $^{22}$Ne and $^{25}$Mg of the order
1\% (e.g. Iben \& Tutukov 1985). Or we see a naked O-Ne-Mg white dwarf which
has undergone carbon burning. A comparison with the respective evolutionary
models (Ritossa et al.\ 1999) suggests that we could look onto such a naked
remnant where $^{20}$Ne and a mixture of several Mg isotopes
($^{24}$Mg,$^{25}$Mg,$^{26}$Mg) should be present. We have already speculated
earlier, that the onset of C burning and subsequent evolutionary processes
might explain the uniqueness of H1504+65. In order to confirm this idea, we
will also look for sodium absorption lines, because the stellar models also
predict a relatively high amount of $^{23}$Na in this case. To conclude,
H1504+65 might have been one of the ``heavyweight'' intermediate-mass stars
(9\,M$_{\odot}\le M \le $ 11\,M$_{\odot}$) which form white dwarfs with
electron-degenerate O-Ne-Mg cores resulting from carbon burning. This
speculation is corroborated by the fact that H1504+65 is one of the most massive
PG1159 stars known (0.86$\pm$0.15\,M$_\odot$).

\begin{acknowledgements}
UV and X-ray data analysis in T\"ubingen is supported by the DLR under grant
50\,OR\,0201. MAB is supported by the Particle Physics and Astronomy Research
Council, UK. JWK is supported by the FUSE project, funded by NASA contract
NAS5-32985.
\end{acknowledgements}

\end{document}